\documentclass[twocolumn,aps,prl,floatfix,10pt,superscriptaddress,longbibliography]{revtex4-2}
{}

\usepackage{xcolor}
\usepackage[utf8]{inputenc} 
\usepackage{bm}
\usepackage{amsmath,amsthm,amssymb}
\usepackage{graphicx} 
\usepackage[
colorlinks=true,
urlcolor=blue,
linkcolor=blue,
citecolor=blue
]{hyperref}
\usepackage[normalem]{ulem}

\def\urlprefix{}
\def\url#1{}

\usepackage{braket}

\definecolor{myblue}{rgb}{0.4, 0.6, 0.902}

\begin{document}
\title{
Interaction induced AC-Stark shift of exciton-polaron resonances
}
\author{T. Uto}
\thanks{These authors contributed equally}
\affiliation{Institute for Quantum Electronics, ETH Z\"{u}rich, CH-8093 Z\"{u}rich, Switzerland}
\affiliation{Institute of Industrial Science, The University of Tokyo,  4-6-1 Komaba, Meguro-ku, Tokyo 153-8505, Japan}
\author{B. Evrard}
\thanks{These authors contributed equally}
\affiliation{Institute for Quantum Electronics, ETH Z\"{u}rich, CH-8093 Z\"{u}rich, Switzerland}
\author{K. Watanabe}
\affiliation{Research Center for Electronic and Optical Materials, NIMS, 1-1 Namiki, Tsukuba 305-0044, Japan}
\author{T. Taniguchi}
\affiliation{Research Center for Electronic and Optical Materials, NIMS, 1-1 Namiki, Tsukuba 305-0044, Japan}
\author{M. Kroner}
\affiliation{Institute for Quantum Electronics, ETH Z\"{u}rich, CH-8093 Z\"{u}rich, Switzerland}
\author{A. {\.I}mamo{\u{g}}lu}
\affiliation{Institute for Quantum Electronics, ETH Z\"{u}rich, CH-8093 Z\"{u}rich, Switzerland}

\begin{abstract}
	Laser induced shift of atomic states due to the AC-Stark effect has played a central role in cold-atom physics and facilitated their emergence as analog quantum simulators. Here, we explore this phenomena in an atomically thin layer of semiconductor MoSe$_2$, which we embedded in a heterostructure enabling charge tunability. Shining an intense pump laser with a small detuning from the material resonances, we generate a large population of virtual collective excitations, and achieve a regime where interactions with this background population is the leading contribution to the AC-Stark shift. Using this technique we study how itinerant charges modify -- and dramatically enhance -- the interactions between optical excitations. In particular, our experiments show that the interaction between attractive polarons could be more than an order of magnitude stronger than those between bare excitons.
\end{abstract}

\maketitle	
\textit{Introduction }---
Atomically thin transition metal dichalcogenides (TMDs) and their van der Waals heterostructures constitute a versatile platform for exploration of phenomena at the frontier of many-body physics and quantum optics \cite{huang2022enhanced,wang2018colloquium,mak2016review}. Arguably, the most significant feature of this new platform is the weak dielectric screening and relatively heavy band-mass of electrons and holes, leading to strong Coulomb interactions and appearance of tightly bound excitons as elementary optical excitations. On the one hand, the small Bohr radius ($a_{\rm ex} \sim 1$nm) of excitons implies strong coupling to light,  which ensures that a pristine monolayer TMD realizes an atomically-thin mirror in the absence of a cavity~\cite{Back2018mirror,scuri2018large} and exhibits large normal-mode splitting between exciton-polariton modes when embedded inside a cavity~\cite{Menon15TMDpolariton,Schneider16TMDpolariton}. On the other hand, electron- or hole-exchange based interaction between two tightly bound excitons is drastically reduced, leading to a predominantly linear optical response~\cite{barachati2018interacting,scuri2018large,tan2020interacting}. Therefore, strong exciton-photon coupling is fundamentally linked to weak exciton-exciton interactions in TMDs, which in turn constitutes a major challenge to the prospect of engineering nonlinear optical devices~\cite{Shahmoon2017,Zeytinolu2017,Ryou2018,Wild2018}.

Different approaches that could meet this challenge by enhancing exciton-exciton interactions without sacrificing strong light-matter coupling have been explored. While promising results are reported~\cite{Cristofolini2012,Togan2018,Rosenberg2018,tan2020interacting}, the large uncertainty in the determination of the underlying exciton-exciton interaction strength has been a major hindrance in assessing and comparing these approaches. Measurement of the interaction-induced blue-shift under direct resonant excitation leads to generation of a sizeable dark exciton population, rendering the extracted interaction strength unreliable. A partial remedy is provided by studying nonlinear response of exciton-polaritons; however, recent theoretical work showed that interactions between exciton-polaritons can be drastically different from those between bare excitons~\cite{christensen2022microscopic}. 

In this Letter, we introduce a novel method to reliably measure exciton-exciton interactions, based on the light shift of the excitonic resonances in response to an intense red-detuned femtosecond laser pulse. Previously, the AC-Stark shift of excitons in TMDs was studied for large pump detuning~\cite{kim2014ultrafast,sie2014}, in a regime well captured by the simple picture of a dressed two-level system, similar to that of a single atom in an off-resonant light field. For very large detunings, comparable to the band gap, the Bloch-Siegert shift becomes significant and has been observed in \cite{sie2017BlochSiegert}.
Here, we are interested in the opposite limit, where the pump detuning from the excitonic resonances is much smaller than the exciton binding energy: in this limit, the pump pulse generates a large population of virtual excitations that exist only during the pump-pulse duration. The interactions between this background of pump-generated virtual excitations and a test excitation produced by the probe pulse provide the dominant contribution to the light shift~\cite{cunningham2019StarkEffect,Slobodeniuk2023,combescot1992semiconductors,schmitt1986collective,zimmermann1990dynamical,haug2009book}.

A possible avenue to enhance exciton-exciton interactions is to embed them in a two dimensional degenerate electron system (2DES). Following seminal studies on III-V quantum wells \cite{Rapaport2000,Rapaport2001,Suris2003}, recent work established that dynamical screening of excitons in a doped TMDs by the 2DES modifies the nature of elementary optical excitations, leading to the formation of attractive- and repulsive-exciton-polarons (AP and RP) \cite{sidler2017fermi,efimkin2017many,huang2023quantum}. Arguably, the principal result of our work is the use of the AC-Stark effect to measure the bare AP interactions as a function of $n_e$, where we demonstrate a dramatic enhancement of the polaron-polaron interaction, up to a factor $\sim35$ as compared to interactions between bare excitons, for $n_e \le 2 \times 10^{11}$~cm$^{-2}$. This behavior was theoretically predicted in \cite{tan2020interacting}, but was not experimentally observed.

\begin{figure}[h!]
	\centering
	\includegraphics[width=\columnwidth]{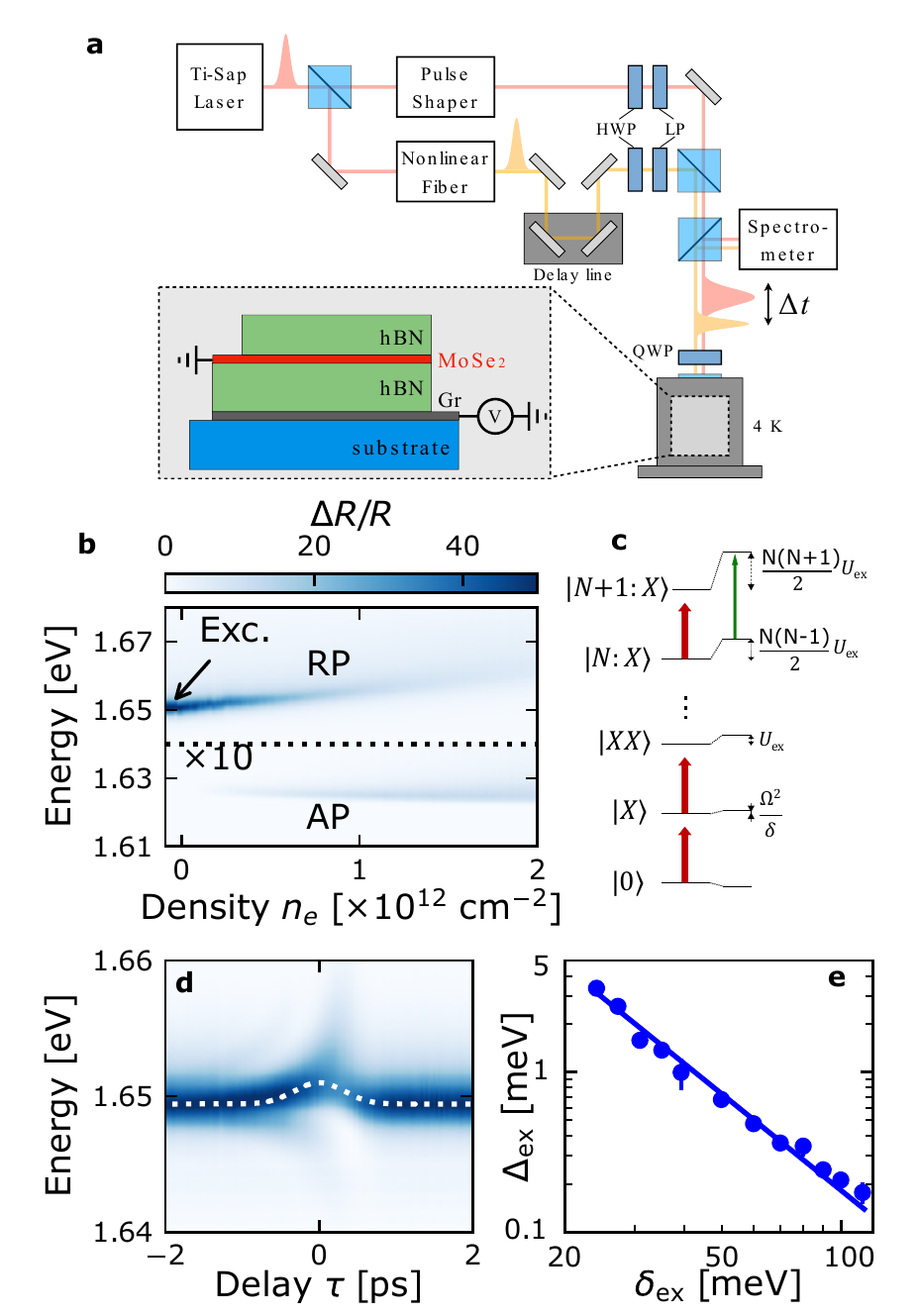}
	\caption{{\bf a} Sketch of the pump-probe setup and the van der Walls heterostructure. {\bf b} Reflection spectrum of the measured device as a function of the electron density $n_e$. In the energy range near the AP resonance ($E<1.64$eV) the reflection signal is multiplied by $10$ to enhance visibility.{\bf c} Schematic of the energy levels showing the usual AC-Stark shift for the first two-levels and the interaction induced light shift which increases with the density of pump-generated excitons. {\bf d} Reflection spectrum at charge neutrality, as a function of the delay between the co-circularly-polarized pump and probe pulses. The white dashed line is a fit from which we extract the amplitude of the light shift at zero time delay. The latter is shown in {\bf e} as a function of the pump-laser detuning from the exciton resonance ($\delta_{\rm ex}$). The blue line is a fit using $\Delta_{\rm ex}=B/\delta_{\rm ex}^2$. For all data presented in the manuscript, the error bars show the statistical error corresponding to two standard deviations obtained from a set of a few repetitions of the experiment.}\label{figure1}
\end{figure}

Our experiments are performed on a monolayer MoSe$_2$ encapsulated in hBN, at cryogenic temperature ($T\lesssim 10\,$K). In these conditions, we achieve narrow exciton lines (with a Lorentzian width of $\approx2.5$\,meV) enabling a near-resonant drive. Our pump-probe setup is sketched in Fig.\,\ref{figure1}\,\textbf{a}. A mode-locked Ti: sapphire laser delivers $\sim100\,$fs pulses at a rate of $76\,$MHz. With a pulse shaper we narrow the bandwidth of the pump (increasing the duration by a factor $\lesssim2$), while a non-linear crystal fiber generates a white-light continuum to probe the exciton and AP transition. Both pulses are focused near the diffraction limit onto the sample using an apochromatic microscope objective (more details about the sample and experimental setup in the Supplementary Material (SM) \cite{SM}).

The evolution of the reflection spectrum as a function of the gate voltage is shown in Fig.\,\ref{figure1}\,\textbf{b}. As charges are introduced, the exciton line smoothly evolves into the RP and a red-shifted AP resonance emerges \cite{sidler2017fermi,efimkin2017many,huang2023quantum}. 

\textit{Exciton light shift }---
To benchmark our method, we first focus on the excitonic light shift at charge neutrality, for a pump laser red-detuned from the exciton resonance and co-circularly polarized with the probe laser. For zero time delay between the two pulses ($\tau = 0$) we observe a blue shift of the exciton line (Fig.~\ref{figure1}\,\textbf{c}), together with a  broadening. The latter stems from the averaging over a spatial and time dependent light shift, since the pump and probe lasers have the same spot size, and comparable duration. For $\tau\lesssim0$ we also observe the emergence of weak sidebands, a common artifact of pump-probe experiments, which can be understood as the free induction decay of probe-generated excitons, perturbed by the pump pulse (for more details see \textit{e.g.} \cite{koch1988transient,haug2009book} or the SM \cite{SM})

For detunings $\delta_{\rm ex}=E_{\rm ex}-E_{\rm pump}$ large compared to the maximum Rabi frequency of the pump laser ($\Omega_{max}$), the light shift can be expanded as \cite{combescot1992semiconductors}
\begin{align}
	\Delta_{\rm ex}\approx\frac{A}{\delta_{\rm ex}}+\frac{B}{\delta_{\rm ex}^2}\,.
\end{align}
Here, the first term corresponds to the usual AC-Stark shift of a two-level system \cite{cohen1977dressed}. The second term arises from many-body effects, namely Coulomb interaction and Pauli blocking, due to the pump-laser-generated population of virtual excitons. These interaction effects are usually described within a Hatree-Fock approximation \cite{schmitt1986collective,zimmermann1990dynamical,haug2009book}, in which case we can write $B/\delta_{\rm ex}^2=U_{\rm ex}n_{\rm ex}$ where $n_{\rm ex}\propto\delta_{\rm ex}^{-2}$ is the exciton density and $U_{\rm ex}$ is an effective exciton-exciton interaction strength. In a regime of intermediate detunings, $\hbar\Gamma_{\rm ex},\Omega_{max} \ll |\delta_{\rm ex}| \ll E_x$, where $\Gamma_{\rm ex}$ is the exciton radiative decay and $E_x$ the exciton binding energy, the interaction-induced light shift is expected to dominate over the single-particle response. This is the regime explored throughout this Letter. Figure\,1\textbf{d} shows that the exciton light shift is indeed well captured by a $1/\delta_{\rm ex}^2$ dependence. From this measurement we extract $U_{\rm ex}\approx 0.09\pm0.03\,\mathrm{\mu eV\mu m^2}$ (for details on the calibration of $n_{\rm ex}$, see the Supplementary Material \cite{SM}). Our estimate is consistent with previous measurements based on two-dimensional spectroscopy \cite{barachati2018interacting}, or on the non-linear resonant response of excitons \cite{scuri2018large} and exciton-polaritons \cite{tan2020interacting}. We point out that all these measurements are an order of magnitude lower than the theoretical expectation, $\sim 3 E_{x}a_{\rm ex}^2\sim 1\,\mathrm{\mu eV\mu m^2}$ \cite{ciuti1998interaction,shahnazaryan2017excitoninteraction}.

For cross-circularly polarized pump and probe lasers, producing excitons in opposite valleys, the dominant contribution to the interaction stemming from hole and electron exchange is suppressed, leaving a negligible direct interaction shift \cite{ciuti1998interaction,shahnazaryan2017excitoninteraction}. On the other hand, the lack of Pauli-blocking allows for the formation of a bound state of two excitons in opposite valleys, called a biexciton. The pump drives the transition from a ``probe exciton" to the biexciton, resulting in an additional contribution to the light shift, which has been previously investigated in \cite{yong2018biexcitonic,sie2016biexciton}. We report similar results in the Supplementary Material \cite{SM}, although we point out that we obtain a biexciton binding energy $E_{\rm binding} = 29\pm1.5$\,meV slightly larger than the values reported in  \cite{hao2017biexciton,yong2018biexcitonic} while being in good agreement with another recent measurement \cite{tan2022bose}. A possible origin of the discrepancy could be the presence of residual charges in devices without electrical gates, screening the Coulomb interaction thereby reducing $E_{\rm binding}$.

\begin{figure}
	\centering
	\includegraphics[width=\columnwidth]{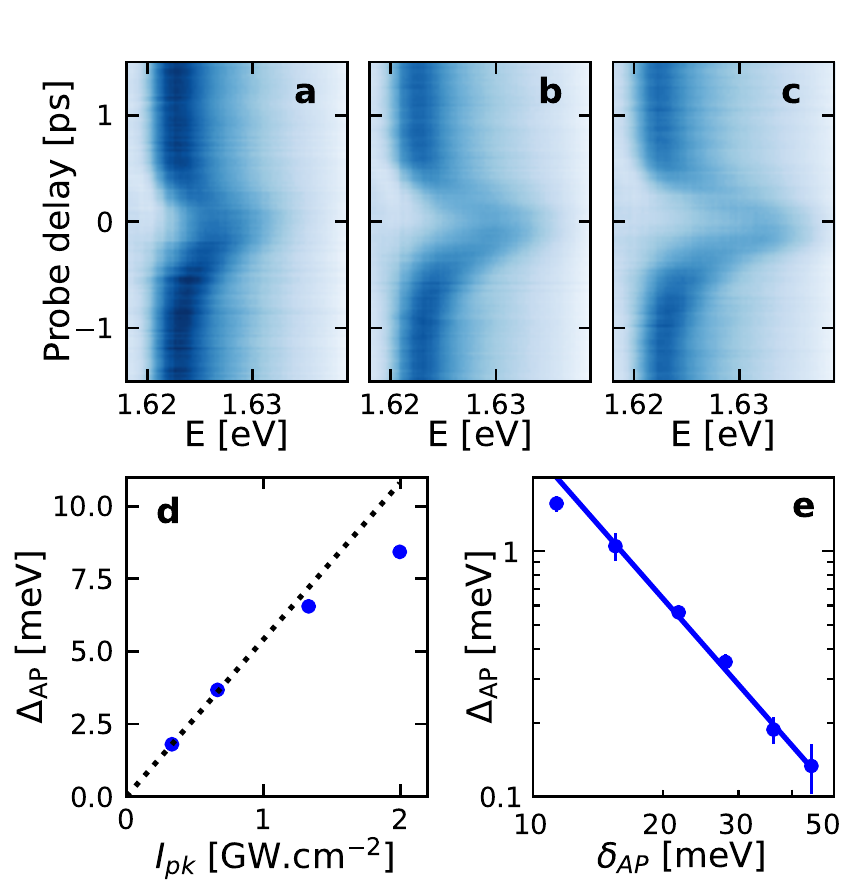}
	\caption{AC-Stark shift of the attractive polaron (AP) resonance for co-circularly polarized pump and probe lasers. In \textbf{a}-\textbf{c}, we show the AP resonance as a function of the pump-probe delay $\tau$ for increasing peak pump laser intensity ($I_{\rm pk}\approx0.7;\,1.3;\,2\,$GW/cm$^2$). The light shift at $\tau = 0$ is plotted in \textbf{d} as a function of the pump intensity, showing deviation from a linear dependence in $I_{\rm pk}$ (dashed line). Here the pump laser detuning from AP is $\delta_{\rm AP}\approx 13\,$meV and the electron density is $n_e\approx1.7\times10^{12}$\,cm$^{-2}$. In {\bf e}, we show the $\delta_{\rm AP}$ dependence of the light shift, which is well fitted by $B/\delta_{\rm AP}^2$, shown as a blue line. Here, $n_e \approx 0.17\times10^{12}$\,cm$^{-2}$ and  $I_{\rm pk}\approx0.4$\,GW\,cm$^{-2}$ so that the $I_{\rm pk}$ dependence is within the linear regime.}\label{figure2a}
\end{figure}

\begin{figure}
	\centering
	\includegraphics[width=\columnwidth]{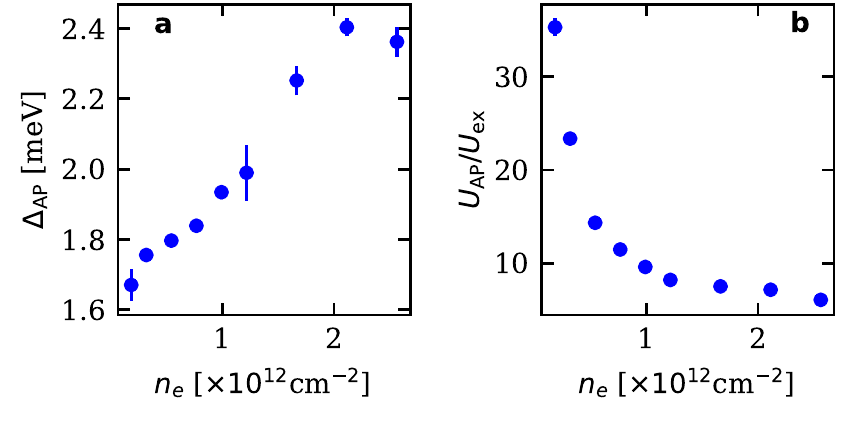}
	\caption{{\bf a} Electron density dependence of the attractive polaron (AP) light shift for co-circularly polarized pump and probe beam. From this data and a measurement of the AP oscillator strength \cite{SM}, the AP-AP interaction strength $U_{\rm AP}$ is extracted and compared to the exciton-exciton interaction strength $U_{\rm ex}$ in {\bf b}. Here, the detuning from the AP resonance is $\delta_{\rm AP}\approx 25\,$meV and the peak pump intensity is $I_{\rm pk}\approx1.7$\,GW\,cm$^{-2}$.}\label{figure2b}
\end{figure}

\begin{figure*}
	\centering
	\includegraphics[width=\textwidth]{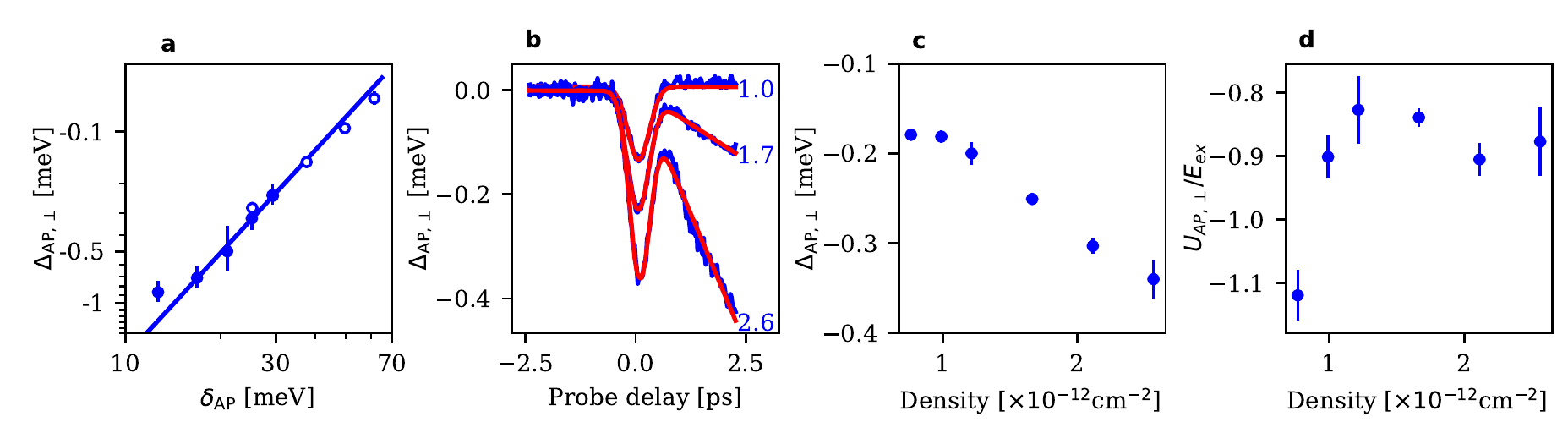}
	\caption{Light shift of the attractive polaron (AP) for cross-circularly polarized pump and probe beam.  \textbf{a} Detuning  $\delta_{\rm AP}$ dependence of the light shift at an electron density of $n_e \approx 1.0\times10^{12}$\,cm$^{-2}$. For large $\delta_{\rm AP}$ (open symbols) we use the full bandwidth of the pump and the peak intensity is $I_{\rm pk}\approx 5.3$\,GW/cm$^{-2}$. To approach the AP resonance, we reduce the bandwidth and consequently $I_{\rm pk}$  by a factor of $\approx0.4$; we then re-scale the data (full symbols), ensuring that the two measurements match for intermediate detunings ($\delta_{AP}\approx25;\,30$meV). The data is well fitted by a $B/\delta_{\rm AP}^2$ law, shown as a blue line. {\bf b} Time dependence of the line shift for various $n_e$ at $\delta_{\rm AP}\approx 25\,$meV. The red line is a fit, from which we extract the shift at $\tau = 0$. The latter is shown in {\bf c} as a function of $n_e$. {\bf d} The ratio of the interaction between opposite valley APs and that of same valley excitons for $I_{\rm pk}\approx1.7$\,GW/cm$^{-2}$.}\label{figure3}
\end{figure*}

\textit{Attractive polaron light shift }---
Having established our approach as a sensitive probe of the interaction between optical excitations, we now turn to the main results of our paper, where we investigate the interactions between APs in an electron doped TMD monolayer. Figure\,\ref{figure2a}\,{\bf a}-{\bf c} shows the AP light shift for co-circularly-polarized pump and probe lasers: despite its relatively low oscillator strength ($f_{\rm AP}$), the large blue shift of the AP well exceeds its linewidth. We remark that the AP resonance is symmetric around $\tau =0$, indicating that the pump laser does not generate incoherent AP population or effects a quench of the 2DES state. This observation should be contrasted with resonant pump-probe experiments carried out using AP-polaritons~\cite{tan2020interacting}. However, for increasing pump intensity $I_{\rm pk}$ we observe in Fig.\,\ref{figure2a}\,{\bf a}-{\bf c} an overall reduction of $f_{\rm AP}$ together with a small red shift ($\approx0.8\,$meV at most), both independent on $\tau$. This latter observation suggests that part of the pump-pulse energy is absorbed by the TMD, which then relaxes on a time scale much longer than the pulse repetition rate. Thereby, depending on the pump intensity, we effectively change the steady-state of the TMD, presumably its temperature and/or charge density - both of which potentially leading to a reduction of $f_{\rm AP}$. 
As a consequence, we observe a sub-linear increase of the light shift with increasing $I_{\rm pk}$ (Fig.\,\ref{figure2a}\,{\bf d}). We point out that the regime of linear dependence of $\Delta_{\rm AP}$ with $I_{\rm pk}$ increases with increasing detuning, and we are able to observe the onset of a saturation in (Fig.\,\ref{figure2a}\,{\bf d}) only because the system is driven close to resonance ($\delta_{\rm AP}\approx13\,$meV). In the following we focus on a range of pump-laser detunings and intensities where the reduction of $f_{\rm AP}$ is negligible and  $\Delta_{\rm AP}\propto I_{\rm pk}$. 

Figure\,\ref{figure2a}\,{\bf d} shows the dependence of the AP light shift on the detuning from the AP resonance ($\delta_{\rm AP}=E_{\rm AP}-E_{\rm pump}$) for $n_e \approx 0.17\times10^{12}$\,cm$^{-2}$. It is well reproduced by a $\Delta_{\rm AP}\propto1/\delta_{\rm AP}^2$ law, which demonstrates that it originates predominantly from AP-AP interactions. We point out that
considering the small $f_{\rm AP}$ as compared to that of the RP oscillator strength, particularly for low $n_e$, one could expect that the pump generates more RP than AP (despite a smaller detuning to the latter). However,  a light shift dominated by the RP population would scale as $\Delta_{\rm AP}\propto n_{\rm RP}\propto1/\delta_{\rm RP}^2$. Even at the lowest electron densities ($n_e\approx 0.17\times10^{12}$cm$^{-2}$) we measured, we do not observe such a detuning dependence and the deviation from $\Delta_{\rm AP}\propto1/\delta_{\rm AP}^2$ law remains negligible \cite{SM}. This observation suggests that the AP-RP interactions are much weaker than the AP-AP interaction, especially for low $n_e$, which is consistent with the fact that the RP has a dominant exciton content in that regime.

Since $f_{\rm AP}$ increases linearly with $n_e$, we would normally expect $\Delta_{\rm AP}$ to also increase linearly with $n_e$ -- indeed, $\Delta_{\rm AP}=U_{\rm AP}n_{\rm AP}$ and $n_{\rm AP}\propto f_{\rm AP} \propto n_e$. At a first glance, this is indeed what we observe in fig.\ref{figure2b}\,{\bf a}. However, we also find that unlike $f_{\rm AP}$, a linear $n_e$ fit to $\Delta_{\rm AP}$ yields a finite value for $n_e = 0$: this striking observation can be explained as an increase of the interaction strength $U_{\rm AP}$ with decreasing $n_e$ that becomes prominent for $n_e \le 2 \times 10^{11}$~cm$^{-2}$, where it counteracts the effect of decreasing $n_{\rm AP}$ or $f_{\rm AP}$. To highlight this feature, we compare the renormalized AP light shift $\Delta_{\rm AP}/f_{\rm AP}$ with that of the exciton light shift $\Delta_{\rm ex}/f_{\rm ex}$, obtained at charge neutrality, for the same pump intensity and at a wavelength such that $\delta_{\rm ex}=\delta_{\rm AP}$. In this way, we obtain the interaction ratio $U_{\rm AP}/U_{\rm ex}=\Delta_{\rm AP}/f_{\rm AP}\times f_{\rm ex}/\Delta_{\rm ex}$  which we plot in Figure\,\ref{figure2b}\,{\bf b}. We observe a dramatic enhancement of the AP-AP interactions -- up to a factor $35$ -- as $n_e$ is lowered.

In a simple but far-reaching ansatz, the AP wave-function is described as a superposition of a zero-momentum bare exciton plus an unperturbed 2DES, and an exciton scattered into a finite momentum state while generating a single particle-hole excitation in the 2DES of the conduction band of the opposite valley \cite{Rapaport2000,Rapaport2001,sidler2017fermi,chevy2006universal,Suris2003}.  The latter contribution could also be considered as a superposition of trion-hole pairs. For low $n_e$, the probability of finding a bare exciton (quasi-particle weight) in an AP excitation is small. Consequently, an AP excitation predominantly generates a collective excitation of tightly bound trions with radius $a_T \sim 2$nm, each surrounded by a Fermi sea hole of extent on the order of the inverse Fermi wavevector $k_F^{-1}$. The depletion of the 2DES around the trion leads to an effective repulsive interaction between two APs, through a partial suppression of hybridization of the bare exciton and collective trion-hole excitations.  The expansion of the depleted region $\propto k_F^{-1} \propto n_e^{-1/2}$ as $n_e$ decreases can thus partially compensate for the reduction of $f_{\rm AP}$ and consequently $n_{\rm AP}$, insuring the persistence of a significant AP light shift for low $n_e$. 

The mechanism outlined above only takes place for APs generated by co-polarized pump and probe lasers \cite{tan2020interacting,muir2022interactionspolaron}. In the cross-polarized configuration, we also observed an AP light shift $\Delta_{\rm AP,\perp}$, albeit much smaller in magnitude and of the opposite sign. Figure\,\ref{figure3}\,{\bf a} shows the detuning dependence of $\Delta_{\rm AP,\perp}$ which is consistent with a $\Delta_{\rm AP,\perp}\propto-1/\delta_{\rm AP}^2$ law, pointing out again to an interaction between the probe and pump laser induced APs. More importantly, we emphasize that our light shift data cannot be fitted with  $\Delta_{\rm AP,\perp}\propto 1/(\delta_{AP}-E_0)$; which could have emerged from coupling to a putative charged biexciton resonance\cite{hao2017biexciton,muir2022interactionspolaron,SM}. 
Figure\,\ref{figure3}\,{\bf b} shows the time dependence of the AP light shift at various densities. For $\tau>0$ (pump before probe), we observe a continuous red shift of the AP line, which increases together with the density $n_e$, possibly due to residual pump-induced high momentum APs. We emphasize that this shift also exists in co-circularly-polarized pump-probe measurements but remains negligible as compared to the AC-Stark shift at $\tau = 0$. To fit the data and extract the coherent response, we use a sum of a Gaussian and a piecewise linear function. Contrary to the co-circularly-polarized case, the amplitude of the Gaussian term (coherent response) increases approximately linearly with the electron density $n_e$ as shown in Fig.\,\ref{figure3}\,{\bf c}; here, we discarded low density $n_e<8\times10^{12}$\,cm$^{-2}$ data for which the fit was unreliable. After proper normalization by $f_{\rm AP}$, we extract the interaction strength between opposite-valley APs, $U_{\rm AP,\perp}$, which we compare to the same-valley exciton interaction $U_{\rm ex}$ in Fig.\,\ref{figure3}\,{\textbf d}:  we observe almost no dependence of $ U_{\rm AP,\perp}$, which remains comparable (in absolute value) to $U_{\rm ex}$ for all $n_e$.

To explain this observation, we consider a $\sigma_-$-polarized pump laser producing APs in the $K'$-valley. The polaron dressing leads to a  $K$-valley electron being promoted to high momentum states with  $k\sim 1/a_T$, thereby reducing the phase-space filling (at low $k$) for a probe-generated $K$-valley AP. For low $n_E$ where  $k_Fa_T\ll 1$ this mechanism could be at the origin of the attractive interactions between opposite-valleys APs. However, further investigations are needed to confirm this hypothesis.

\textit{Conclusion and Outlook }---
Our work establishes the AC-Stark effect as a novel approach to measure the interaction between optical excitations in charge tunable TMDs monolayers. An extension of this technique to assess the modification of interactions due to the formation of moiré heterostructures in (twisted) heterobilayers presents no difficulties. 
By using a detuned pump laser, we generate a large virtual exciton or AP population and thereby almost fully suppress dark-exciton generation, which plagued previous studies \cite{tan2020interacting}. 
An exciting application of the technique we developed would rely on a Laguerre-Gauss pump beam to shift away the AP resonance, except in a small region around the beam's vortex. Increasing the pump laser strength could reduce the region where the AP resonance does not experience a blue shift down to a size of $\sim1/k_F$; in this limit, we could use strong AP-AP interactions to generate sub-Poissonian light. For that purpose, low electron densities (small $k_F$) are preferable, and our finding of a persistent AP light shift as we lower $n_e$ is thus of paramount importance.

The data that support the findings of this Letter are
available in the ETH Research Collection \cite{DOIData}.

We are grateful to Andrea Bergschneider, Alperen Tugen, and Francesco Colangelo for their contribution to the design and fabrication of the optical setup and of the device. We also thank Feng Wang for insightful discussions. This work was supported by the Swiss National Science Foundation (SNSF) under Grant Number 200020$\_$207520. B\,.E. acknowledges funding from an ETH postdoc fellowship.

\newpage
\clearpage
\renewcommand{\thefigure}{S\arabic{figure}}
\setcounter{figure}{0}    

\onecolumngrid

	{\Large \begin{center}
			Supplementary material
		\end{center}}
	\vspace{1cm}
	\twocolumngrid
	
	\section{Device and electron density estimation}\label{Sec: Device}
	Our device structure is sketch in Fig.\,1 of the main text. It consists of a monolayer MoSe$_2$ encapsulated in hBN flakes with a thicknesses estimated from their optical contrast to be $d_{\rm top}\approx 21\pm3\,$nm (top layer) and $d_{\rm bot}\approx 42\pm5\,$nm (bottom layer). Underneath the lower hBN, a few-layer graphite flake is used as a gate to electrostatically dope the sample. All flakes were mechanically exfoliated from bulk crystals, and stacked  using a standard dry-transfer
	technique \cite{zomer2014fast} with a poly(bisphenol A carbonate) film on a
	polydimethylsiloxane (PDMS) stamp. The heterostructure was then deposited on a Si/SiO$_2$ (285 nm) substrate, and the graphite and MoSe$_2$ flakes were then contacted using optical lithography and electron beam metal deposition.
	
	To introduce electrons in the TMD, we ground it while applying a voltage $V$ to the graphite gate. Neglecting the quantum capacitance of the doped TMD monolayer, we obtain the charge density from the relation $n_{e}e=C(V-V_0)$, where $e$ is the electron charge, $V_0$ is the gate voltage at the onset of doping and $C=\epsilon_0\epsilon_{\rm hBN}^\perp/d_{\rm bot}$ is the geometric capacitance. For the hBN out-of-plane dielectric constant we use the value $\epsilon_{\rm hBN}^\perp=3.5$ \cite{laturia2018dielectric,kim2012synthesis}. The assumption of a negligible quantum capacitance has been verified in previous studies of Van der Waals heterostructures, where the charge density could be infered e.g. from the filling of a Moiré superlattice\,\cite{Regan2020,Tang2020,Shimazaki2020} or of Landau levels\,\cite{TomaszLL}. It is further supported in our sample by the observation of a linear dependence  of the AP oscillator strength $f_{AP}\propto (V-V_0)$ at low density (See Sec.\,\ref{Sec: Reflection data}).

	\section{Experimental setup}
\begin{figure}
		\centering
		\includegraphics[width=\columnwidth]{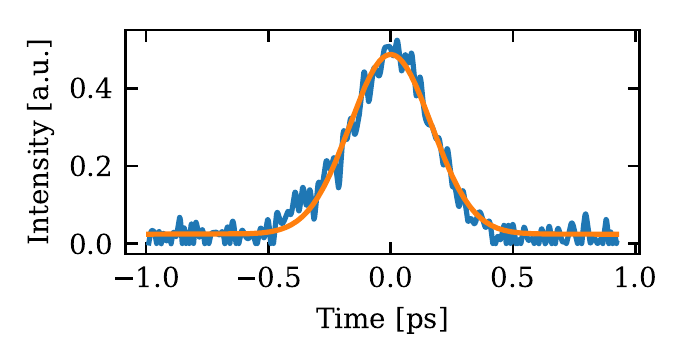}
		\caption{Time profile of the pump pulse measured using an autorcorrelator. From a Gaussian fit, we obtain a duration of $\approx 270\,$fs (FWHM).}\label{figurePumpPulse}
	\end{figure}

	The sample is loaded in a dry cryostat (attodry800) with free space optical access.
	Our main light source is a pulsed Ti:sapphire laser (Tsunami, Spectra-physics), with a repetition rate of $76\,$MHz and a pulse duration of $\approx 140\,$fs (FWHM, before pulse-shaping). The pulses are split into the pump and probe paths, before being recombined, as shown in Fig.\,1 of the main text. For the pump beam, we reduce the bandwidth using a pulse-shaper, typically to $\approx 5\,$meV, resulting in a duration of $\approx 270\,$fs. We show a typical time profile obtained using an autocorrelator in Fig.\,\ref{figurePumpPulse}, from which we can extract the pulse duration using a Gaussian fit.
	
	For the probe pulse, we use a non-linear crystal fiber (femtowhite 800, NKT photonics) to generate a broad continuum spanning the exciton/polarons resonances. Both pulses are focused near the diffraction limit onto the sample using a microscope objective with NA$\approx0.8$ (LT-APO/VISIR/0.82). We typically use average power of $\sim 10^2\mu$W for the pump and below microwatt for the probe. For the pump, this results in peak intensity of order $\sim 1$ GW/cm$^2$, and fluences of order $\sim 10^2$uJ/cm$^2$.
	
	\section{Estimation of the exciton density and interaction strength}
	\begin{figure}
		\centering
		\includegraphics[width=\columnwidth]{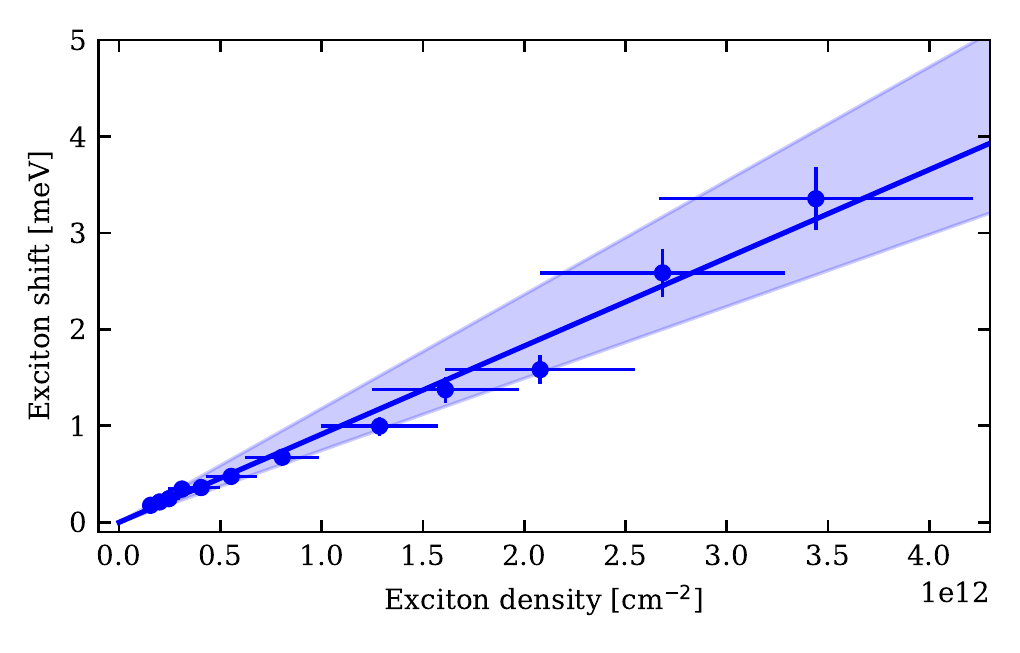}
		\caption{Exciton energy shift as a function of the density of excitons generated by the pump beam. The $y$ error bars correspond to the statistical error estimated by repeating the measurement $3$ times. The $x$ error bars correspond to the systematic error arising from the uncertainty on the various parameters (in particular the exciton (non)radiative decay rate) which are used to estimate the exciton density. The blue line is a linear fit (with the confidence interval shown as a blue cone), from which we extract the interaction strength $U_{\rm ex}\approx 0.09\pm0.03\,\mathrm{\mu eV\mu m^2}$.}\label{SMFigure1}
	\end{figure}
	
	In order to estimate the exciton-exciton interaction strength, we need to evaluate the density of excitons $n_{\rm ex}$ generated by the pump beam. From the optical Bloch equation, we can relate $n_{\rm ex}$ to the flux of photon $I_{\rm i}$ impinging on the TMD \cite{scuri2018large} 
	\begin{align}
		n_{\rm ex}=\frac{2I_{\rm i}R_0}{\gamma_r(1+\delta_{\rm ex}^2/\tilde{\gamma}^2)}\,,
	\end{align}
	where $\gamma_{\rm r}$ ($\gamma_{\rm nr}$) is the (non-)radiative decay rate, $\tilde{\gamma}=(\gamma_{\rm nr}+\gamma_{\rm r})/2$ is the dephasing rate and $R_0=\gamma_{\rm r}^2/(\gamma_{\rm r}+\gamma_{\rm nr})^2$ the TMD reflectance on resonance. From a transfer matrix simulation \cite{scuri2018large} we obtain $\gamma_{\rm r} = 2.4\pm0.4$\,meV, $\gamma_{\rm nr} = 0.6\pm0.3\,$meV, hence $R_0 = 0.65\pm0.13$ and we can deduce the field on the TMD (and hence $I_{\rm i}$) as a function of the incoming field. The error bar comes from the uncertainty on the hBN thicknesses and refractive index (see Sec.\,\ref{Sec: Device}) which are inputs of the TMM simulation.
	
	We show in Fig.\,\ref{SMFigure1} the measured exciton light shift $\Delta$ as a function of the exciton density. In this measurement the light intensity is kept to a fixed value $I_{\rm peak}\approx1.7\pm0.3\,$GW/cm$^2$ ($\approx 170\,$uW average power) and we scan the detuning $\delta$ in a range $30 - 110\,$meV. We observe a linear dependence $\Delta = U_{\rm ex}n_{\rm ex}$ and extract the interaction strength from a fit, $U_{\rm ex}\approx 0.09\pm0.03\,\mathrm{\mu eV\mu m^2}$.
	
	\section{Fitting of the reflection data}\label{Sec: Reflection data}
\begin{figure}
		\centering
		\includegraphics[width=\columnwidth]{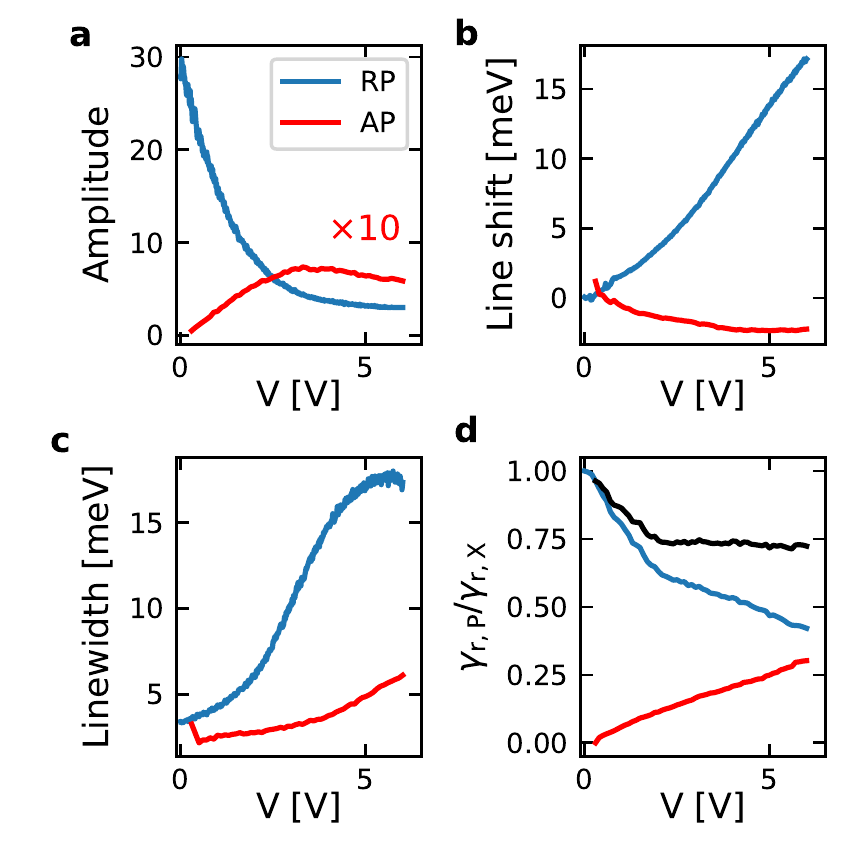}
		\caption{Results of the fitting of the AP and RP reflection spectrum. We use the fitting function (\ref{eq. fitting function}). In {\bf a}, {\bf b}, {\bf c}, we show the fitted amplitude $A(V)$ (multiplied by ten for the AP), resonance shift $E_0(V)-E_0(0)$ and linewidth $\sigma(V)$ as a function of the gate voltage $V$ for the AP (red line) and exciton/RP resonance (blue line). The onset of doping occurs at $V_0\approx0.35\,$V. The last panel {\bf d} show $\gamma_{\rm r,AP}/\gamma_{\rm r,ex}$ (red) and $\gamma_{\rm r,RP}/\gamma_{\rm r,ex}$ (blue), the radiative decay rate of the AP and RP, normalized to that of the exciton. The black line is the sum $(\gamma_{\rm r,AP}+\gamma_{\rm r,RP})/\gamma_{\rm r,ex}$, showing a small decay at low density}\label{FigureFitRes}
	\end{figure}

	Due to the interference of the light reflected at the different dielectric interfaces of our stack, our reflection spectrum are fitted by a sum of a Lorentzian and a dispersive Lorentzian
	\begin{align}
		S(E)=A\left[\cos\theta\frac{\frac{\sigma^2}{2}}{(E-E_0)^2+\frac{\sigma^2}{4}}+\sin\theta\frac{\sigma(E-E_0)}{(E-E_0)^2+\frac{\sigma^2}{4}}\right]\,.\label{eq. fitting function}
	\end{align}
	Here $E$ is the reflected photon energy, and $A$, $\sigma$, $E_0$ are fitting parameters which depend on $n_{\rm e}$. The last fitting
	parameter $\theta$ is independent on $n_e$; in practice, we let it
	free and verify that we obtain the same values for all fits.
	We always fit independently the AP and RP resonance.  We show in Fig.\,\ref{FigureFitRes} the results of the fit as a function of the gate voltage. 
	
	In addition we also fit the data with the reflection spectrum predicted by transfer matrix simulation \cite{scuri2018large}, with the AP/RP energy and (non)radiative decay rates taken as fitting parameter. From this analysis, we obtain the ratio of the oscillator strength $f_{\rm AP,RP}/f_{\rm ex}=\gamma_{\rm r,AP/RP}/\gamma_{\rm r,ex}$, shown in Fig.\,\ref{FigureFitRes}\,{\bf d}, which we then use to infer the interaction strength ratio $U_{\rm AP}/U_{\rm ex}$ from a measurement of the light shift as described in the main text.
	
	\section{Coupling to the biexciton state}\label{sec. biexciton}
	\begin{figure}
		\centering
		\includegraphics[width=\columnwidth]{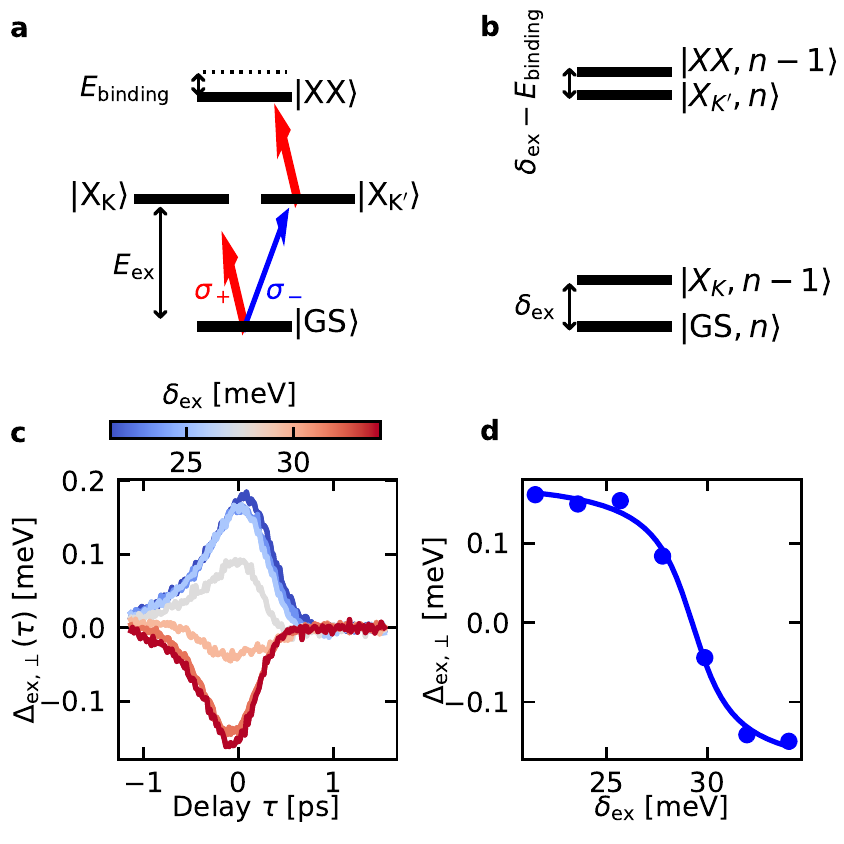}
		\caption{Exciton light shift in cross-polarization. The relevant energy levels are shown in {\bf a}. The $\sigma_+$-polarized pump laser (red arrow) drives the transition between the biexciton state $\ket{XX}$ and the $K'$-valley exciton $\ket{X_{K'}}$ produced by the $\sigma_-$-polarized probe laser (blue). Near the resonance $\delta_{ex}\approx E_{\rm binding}$, the dominant contribution to the light shift of the $K'$-valley exciton comes from level repulsion between the dressed state $\ket{XX,n-1}$ and $\ket{X_{K'},n}$, as shown in {\bf b}. This results in a sign change of the light shift $\Delta_{\rm ex,\perp}$, plotted as a function of the pump-probe delay $\tau$ in {\bf c}. In {\bf d}, we show the light shift at zero time-delay as a function of the pump detuning, which is fitted (solid blue line) to extract the biexciton binding energy $E_{\rm binding}$. }\label{FigureBiexciton}
	\end{figure}
	
	In this section, we report on our observation of the exciton light shift (at charge neutrality), for cross-circularly polarized pump and probe laser. In this situation, which has been studied in \cite{hao2017biexciton,yong2018biexcitonic}, a new contribution to the light shift arise from the coupling to a biexciton sate, a bound state of two excitons in opposit valleys. The energy level structure to consider is shown in Fig.\,\ref{FigureBiexciton}\,{\bf a} and in the dressed state basis in {\bf b}. When the pump laser is detuned from the exciton transition by the biexciton binding energy  $\delta_{\rm ex}=E_{\rm binding}$, we expect an avoided crossing between the dressed states $\ket{X,n}$ and $\ket{XX,n-1}$, where $X$ and $XX$ respectively designate the exciton and biexciton states, and the integers $n$, $n-1$ corresponds to the number of pump photons \cite{cohen1977dressed}. 
	Indeed, we observe a change of sign of the light shift as we sweep the pump-laser detuning across the biexciton resonance $\delta_{\rm ex}=E_{\rm binding}$, as shown in Fig.\,\ref{FigureBiexciton}\,{\bf c}\,,{\bf d}. The amplitude of the shift always remain much smaller than the exciton linewidth and decrease near resonance, where we mostly observe a broadening of the transition and did not resolve the expected Autler-Towns splitting. From a heuristic fit $\Delta_{\rm ex,\perp}=a\arctan[(\delta_{\rm ex}-E_{\rm binding})/b]+c$\,, of the light shift at zero time delay (panel {\bf c}), we extract the biexciton binding energy $E_{\rm binding} = 29\pm1.5$\,meV. As mention in the main text, this value is slightly larger than the estimates of \cite{hao2017biexciton,yong2018biexcitonic}, which could be due to the presence of residual charges in devices without electrical gates, screening the Coulomb interaction and thereby reducing the biexciton binding energy.

	\section{Fit of the light shift wavelength dependence}
	In this section we provide more details on the fitting of the light shift $\Delta_{\rm AP}$ as a function of the pump detuning. We focus on the analysis of the AP, but the data for the exciton in co-polarization is similar to that of the AP, while the case of cross-polarization is described in Sec.\,\ref{sec. biexciton}.
	
	Quite generally, from second order perturbation theory, the light shift can be expanded as \cite{combescot1992semiconductors}
	\begin{align}
		\Delta_{\rm AP}=\frac{A}{\delta_{\rm AP}}+\frac{B}{\delta_{\rm AP}^2}+\frac{C}{\delta_{\rm RP}^2}+\frac{D}{\delta_{\rm AP}-E_{\rm binding}}\,,\label{eq: DetuningDependance}
	\end{align}
	where the first term corresponds to the usual ac-Stark shift arising from light-matter dressing, the second and third terms respectively capture the AP-AP and AP-RP interaction, and the last term can arise from a coupling to a charged biexciton. \\
	For the small detunings (compared to the exciton Rydberg) that we investigate, we expect the interaction terms to dominate over the light-matter dressing. Furthermore, we also expect the AP-AP interaction to play a dominant role compared to the AP-RP interaction. Indeed, our pump laser being red-detuned from the AP resonance, we always have $\delta_{\rm AP}<\delta_{RP}$, and we typically have $n_{\rm AP}\ll n_{\rm RP}$. This strong inequality is however not always satisfied, in particular at low density when $f_{\rm AP}\ll f_{\rm RP}$. On the other hand, this regime also corresponds to the situation where the exciton content is very large for the RP and conversely very small for the AP, resulting in $U_{\rm AP-AP}\gg U_{\rm AP-RP}$.  We therefore always expect $C/\delta_{\rm AP}^2=U_{\rm AP-AP}n_{\rm AP}\gg D/\delta_{\rm RP}^2=U_{\rm AP-RP}n_{\rm RP}$. Finally, same-valley excitons are not expected to form a stable bound state due to Pauli blocking, and therefore we do not expect the last contribution in (\ref{eq: DetuningDependance}) to arise for co-circular pump and probe polarization. For cross-circular polarization, at charge neutrality a biexciton exists and we report on his contribution to the AC-Stark shift in Sec.\,\ref{sec. biexciton}. In the presence of electrons, charged biexciton have been observed in WS$_2$ \cite{muir2022interactionspolaron}. However, Pauli blocking is inhibiting the formation of a charged biexcion for Molybdenum compounds \cite{hao2017biexciton}, where the resident electron and that of the exciton occupy the same lower conduction band (contrary to Tungsten compounds, where bright excitons have their electron in the upper conduction band).
	
\begin{figure}
		\centering
		\includegraphics[width=\columnwidth]{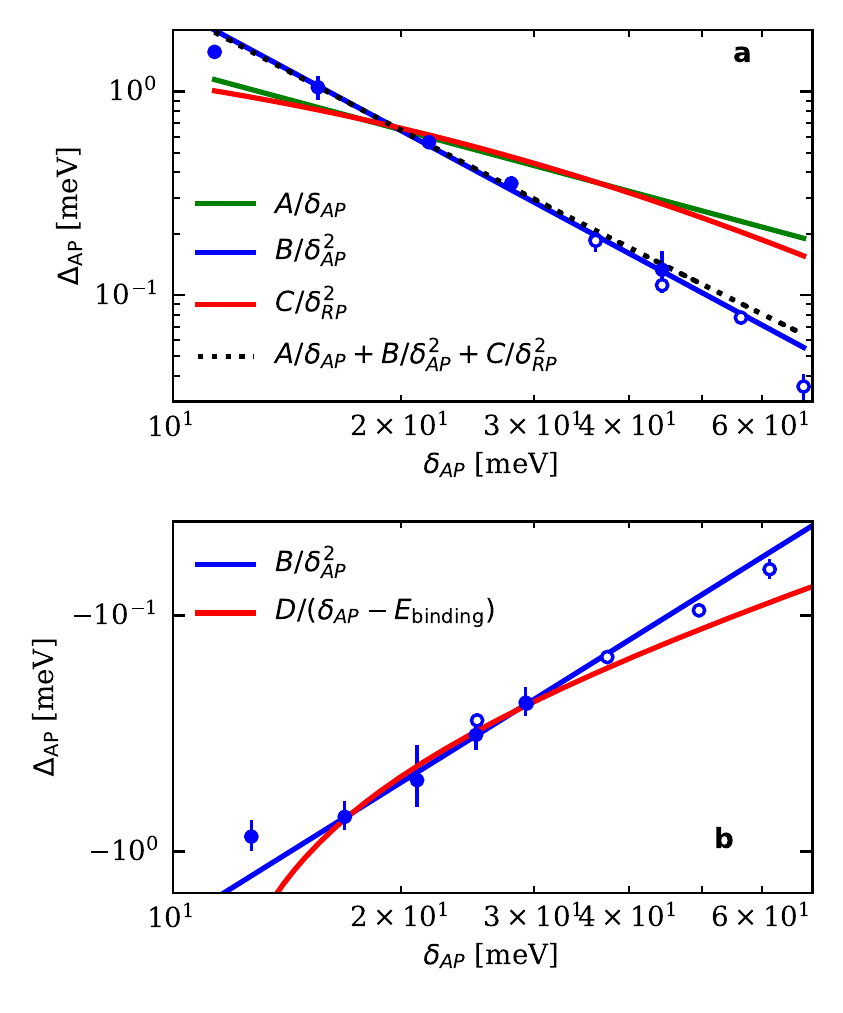}
		\caption{Fits of the wavelength dependence of the AP light shift, for co- ({\bf a}) and cross- ({\bf b}) circularly pump-probe polarization. In both panel, we combine two data sets, one obtain with low pump intensities (at small detuning, blue filled data points) and one obtained at larger pump intensities (at larger detuning, white filled data points). The latter are then rescaled by the intensity ratio. In this way, we are able to keep a decent signal-to-noise ratio as we increase the detuning, while remaining in the linear-in-intensity regime closer to resonance.}\label{FigureFits}
	\end{figure}

	From all these considerations, we expect the AP light shift to follow a law $\Delta_{\rm AP}=B/\delta_{\rm AP}^2$ in the regimes explored in our work. As shown in Fig.\,\ref{FigureFits}, we can recover this result from a fit to our data, without prior knowledge. 
	
	First, in panel {\bf a}, we show the case of co-circularly polarized pump-probe. A one-parameter fit $B/\delta_{\rm AP}^2$ captures very well our measurements; the largest deviation from the fit, observed at small detuning, could be due to absorption. In contrast our data cannot be reproduced by a fit in $A/\delta_{\rm AP}$ or $C/\delta_{\rm RP}^2$. Finally, in a three parameter fit including these three terms, the AP-AP interaction terms contributes to at least $93\%$ of the light shift (at the largest detuning).
	
	We observe a similar behavior for cross-polarization (up to a sign change of $B$). In that case, we also attempt a two-parameters fit $D/(\delta_{\rm AP}-E_{\rm binding})$, which yields a poor agreement with the data, thereby excluding a possible shift induced by the transition to a charged biexciton state.

	\section{Simulation of the line distortion}
\begin{figure}
		\centering
		\includegraphics[]{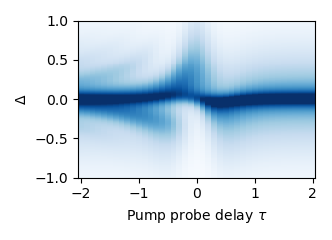}
		\caption{Simulation of the pump-probe experiments taking into account the finite duration of the laser pulses, which results in a broadening of the resonance and side bands at negative time delay.}\label{fig. SimuPumpProbe}
	\end{figure}

	It can be seen in figure\,1\,{\bf c} of the main text that the effect of the pump on the exciton resonance is more complex than a bare line shift. We observe (i) a broadening of the resonance and (ii) at negative time delay, the appearance of side peaks, both on the blue and red side of the exciton resonance. In an adiabatic approximation, We have an instantaneous and local light shift $\Delta(x,t)\propto I_{\rm pump}(x,t)$. The duration of the pump pulse is about twice that of the probe pulse, and both beams are focused to the same spot size. These two effects results in an inhomogeneous (in both time and space) light shift, which readily explains (i). As mentioned in the main text, the fact that the resonance almost perfectly recovers excludes non-coherent effects, such as heating or photo doping, as significant broadening mechanisms. The appearance of side peaks (ii) is a more complex but well known artifact of pump-probe experiments, which is studied in details in e.g. \cite{koch1988transient,haug2009book}. Briefly, the side peaks can be understood as the free induction decay of a real population of exciton injected by the probe and perturbed by the pump - which explains why this effects shows up when the probe comes before the pump, i.e. for $\tau\lesssim 0$. A quantitative treatment requires the resolution of the semiconductor optical Bloch equation beyond the adiabatic approximation. Here, we propose a very simple approach, which can capture the appearance of both blue and red detuned peaks, but which is not meant to be quantitatively accurate. We consider that the polarization generated by the probe pulse read
	\begin{align}
		P(t)=&\cos\left[(\omega+\Delta_0g_{\rm pump}(t)^2)t\right]\times\nonumber\\
		&P_0\left[g_{\rm probe}(t-\tau)+P_1\Theta(t-\tau)\mathrm{e}^{-\gamma (t-\tau)}\right]\,\label{eq: ProbePolarization}
	\end{align}
	where $g_{\rm pump}$ ($g_{\rm probe}$) is the envelope of the pump (probe) electric field, which we take to be the hyperbolic secant function with width $\sigma_{\rm pump}$ ($\sigma_{\rm probe}$). We considered that the pump (probe) pulse arrives at $t=0$ ($t=\tau$). The free induction decay responsible for the side peaks corresponds to the second term in (\ref{eq: ProbePolarization}), where $\Theta$ is a step function (we take the hyperbolic tangent with width $\sigma_{\rm probe}$) and $\gamma$ corresponds to the exciton decay rate. We Fourier transform $P(t)$ and shows its modulus square in Fig.\,\ref{fig. SimuPumpProbe}, for $\omega=2400\,$rad.ps$^{-1}$; $\Delta_0/\omega=2\times10^{-3}$; $\sigma_{\rm pump}=0.4\,$ps; $\sigma_{\rm probe}=0.1\,$ps; $P_1/P_0=0.1$; $\gamma = 1.4\,$ps$^{-1}$. These parameters are not determine after a precise calibration, there chosen are in a realistic range to observe clear side peaks, as can be seen in Fig.\,\ref{eq: ProbePolarization}. In particular we have chosen $\sigma_{\rm probe}/\sigma_{\rm pump}$ slightly smaller than in the actual experiments, for the purpose of having brighter side peaks. 
	
	To conclude we point out that these artifact are more pronounced for the exciton resonance, compared of the AP. This could be due to a smaller ratio $P_1/P_0$ for the AP.
	
	\section{Repulsive polaron light shift}
	\begin{figure*}
		\centering
		\includegraphics[width=\textwidth]{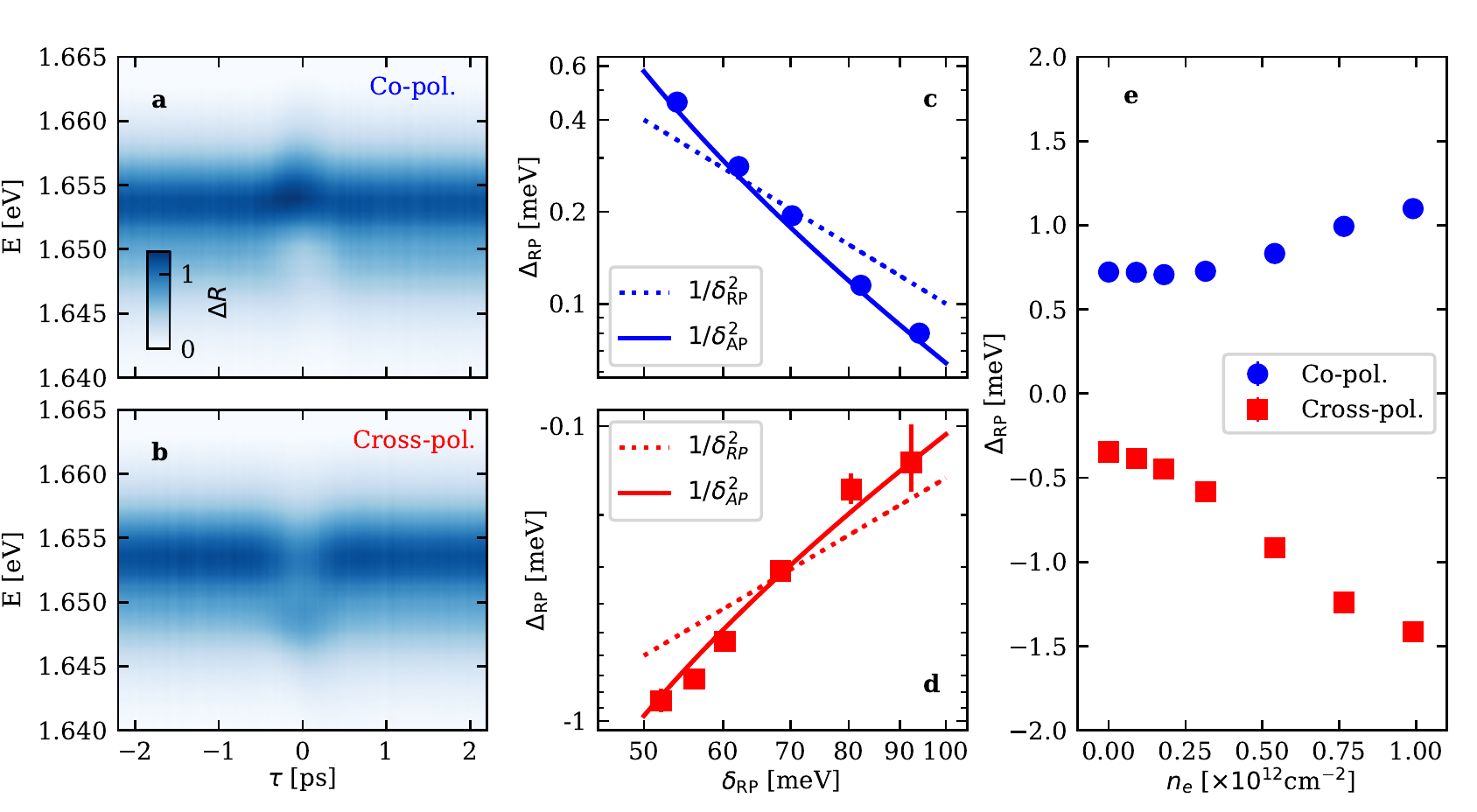}
		\caption{{\bf Repulsive polaron light shift}. Evolution of the reflection spectrum as a function of the pump-probe delay, in co- ({\bf a}) and cross-circular polarization ({\bf b}). At these relatively large electronic density $n_e\approx10^{12}$cm$^{-2}$, the detuning dependance of the light shift ({\bf c} and {\bf d}) is better fitted by a $1/\delta_{AP}^2$ instead of $1/\delta_{RP}^2$. This observation indicates that AP-RP interactions are the leading contribution to the light shift in the detuning range explored here. This effect, together with an overall increase of the interaction due to the polaronic dressing, contributes to an increase of the light shift with the electronic density ({\bf e}).
			These data were obtained in the same experimental condition as for the AP light shift shown in Fig.\,3 and 4 of the main text, i.e. with  $I_{\rm pk}\approx1.7$\,GW\,cm$^{-2}$ and $\delta_{\rm AP}\approx 25\,$meV, $\delta_{\rm RP}\approx 50\,$meV, (note that these are the detuning at $n_e\to0$, they slightly change with increasing $n_e$, see Fig.\,\ref{FigureFitRes}\,{\bf b}).}\label{FigureRP}
	\end{figure*}

	In the main text we focused our analysis of the AC-Stark shift of the attractive polaron. Here we breifly discuss the case of the repulsive polaron (RP). 
	
	Similarly to the exciton or AP, we observe a blue shift of the RP for co-circularly polarized pump and probe pulse, and a red shift for cross-circular polarization, see Fig.\,\ref{FigureRP}\,{\bf a},{\bf b}. The former is driven by repulsive interaction, as described for the exciton or AP. The latter can have two origins, the coupling to the biexciton (discussed in Sec.\,\ref{sec. biexciton}), in particular at low densities when the exciton content of the RP is large and attractive interaction (as observed for the AP) at larger densities when the polaronic dressing is more prominent. In that second situation, due to the transfer of oscillator strength from the RP to the AP for increasing $n_e$ (see Fig.\,\ref{FigureFitRes}), both RP-RP and AP-RP interactions should be taken into account:
	\begin{align}
		\Delta_{\rm RP}=U_{\rm RP-RP}n_{\rm RP}+U_{\rm AP-RP}n_{\rm AP}\,.
	\end{align}
	The role of AP-RP interaction is further enhanced for a red-detuned pump, such that $\delta_{\rm AP}<\delta_{\rm RP}$. The relative contribution of the RP-RP and AP-RP interaction can be deduced from the dependence of the light shift on the pump detuning. We observe in Fig.\,\ref{FigureRP}\,{\bf c}\,{\bf d}, that for both co- and cross-polarization, the RP light shift is indeed better fitted by a $1/\delta_{AP}^2\propto n_{\rm AP}$ law instead of $1/\delta_{RP}^2\propto n_{\rm RP}$, showing that at such electronic density $n_e\approx 10^{12}$cm$^{-2}$, the AP-RP interactions are the leading contribution to the light shift. 
	For increasing electronic density, the enhancement of the interaction  counter-acts the reduction of the oscillator strength and we observe an increase of the light shift in both polarization (Fig.\,\ref{FigureRP}\,{\bf e}). Note that we restricted our analysis to densities $n_e\lessapprox10^{12}\,$cm$^{-2}$ as the RP resonance becomes too broad for larger densities and the fit unreliable.

\bibliographystyle{apsrev4-1}
\bibliography{LightshiftBiblio}

\begin{thebibliography}{51}
\bibitem [{SM()}]{SM}%
  \BibitemOpen
  \href@noop {} {}\bibinfo {howpublished} {for more details see {S}upplemental
  {M}aterial, which includes Refs [45-51]}\BibitemShut {NoStop}%
\end{thebibliography}

\end{document}